\documentclass[fleqn,usenatbib]{mnras}

\usepackage[T1]{fontenc}
\usepackage{ae,aecompl}

\usepackage{graphicx}	
\usepackage{amsmath}	
\usepackage{amssymb}	
\graphicspath{images/}
\usepackage{xcolor}
\usepackage{float}
\usepackage{sidecap}

\newcommand{\cps}{\,cts\,s$^{-1}$}	
\newcommand{\csim}{$\sim$\,}

\title[Type-I X-ray Bursts from MAXI~J1807+132]{Discovery of Thermonuclear Type-I X-ray Bursts from the X-ray binary  MAXI~J1807+132}

\author[A. C. Albayati et al.]{
A.~C.~Albayati,$^{1}$\thanks{E-mail: a.c.albayati@soton.ac.uk}
D.~Altamirano,$^{1}$
G.~K.~Jaisawal,$^{2}$
P.~Bult,$^{3,4}$ 
S.~Rapisarda,$^{5}$ \newauthor
\hspace{0.15cm}G.~C.~Mancuso,$^{6,7}$
T.~G\"uver,$^{8,9}$
Z.~Arzoumanian,$^{4}$ 
D.~Chakrabarty,$^{10}$\newauthor
\hspace{0.15cm}J.~Chenevez,$^{2}$
J.~M.~C.~Court,$^{11}$
K.~C.~Gendreau,$^{4}$
S.~Guillot,$^{12,13}$
L.~Keek,$^{14}$\newauthor
\hspace{0.15cm}C.~Malacaria,$^{15,16}$
and T.~E.~Strohmayer$^{4,17}$ 
\vspace{0.2cm}
\\
$^{1}$School of Physics and Astronomy, University of Southampton, Southampton, SO17 1BJ, UK\\ 
$^{2}$National Space Institute, Technical University of Denmark, Elektrovej 327-328, DK-2800 Lyngby, Denmark\\ 
$^{3}$Department of Astronomy, University of Maryland, College Park, MD 20742, USA\\
$^{4}$Astrophysics Science Division, NASA's Goddard Space Flight Center, Greenbelt, MD 20771, USA\\ 
$^{5}$Shanghai Astronomical Observatory, Chinese Academy of Sciences, 80 Nandan Road, Shanghai 200030, China\\ 
$^{6}$Instituto Argentino de Radioastronom\'{\i}a (CCT-La Plata, CONICET; CICPBA), C.C. No. 5, 1894 Villa Elisa, Argentina\\
$^{7}$Facultad de Ciencias Astron\'omicas y Geof\'{\i}sicas, Universidad Nacional de La Plata, Paseo del Bosque s/n, 1900 La Plata, Argentina\\
$^{8}$Istanbul University, Science Faculty, Department of Astronomy and Space Sciences, Beyaz\i t, 34119, Istanbul, Turkey\\
$^{9}$Istanbul University Observatory Research and Application Center, Istanbul University 34119, Istanbul Turkey\\
$^{10}$MIT Kavli Institute for Astrophysics and Space Research, Massachusetts Institute of Technology, Cambridge, MA 02139, USA\\
$^{11}$Department of Physics and Astronomy, Texas Tech University, PO Box 41051, Lubbock, TX 79409, USA\\ 
$^{12}$CNRS, IRAP, 9 avenue du Colonel Roche, BP 44346, F-31028 Toulouse Cedex 4, France\\
$^{13}$Universit\'{e} de Toulouse, CNES, UPS-OMP, F-31028 Toulouse, France\\
$^{14}$cosine B.V., Oosteinde 36, 2361 HE Warmond, The Netherlands\\
$^{15}$NASA Marshall Space Flight Center, NSSTC, 320 Sparkman Drive, Huntsville, AL 35805, USA\\
$^{16}$Universities Space Research Association, Science and Technology Institute, 320 Sparkman Drive, Huntsville, AL 35805, USA\\
$^{17}$Joint Space-Science Institute, NASA's Goddard Space Flight Center, Greenbelt, MD 20771, USA
}

\date{Accepted XXX. Received YYY; in original form ZZZ}

\pubyear{2020}

\begin{document}
\label{firstpage}
\pagerange{\pageref{firstpage}--\pageref{lastpage}}
\maketitle

\begin{abstract}

MAXI~J1807+132 is a low-mass X-ray binary (LMXB) first detected in outburst in 2017. Observations during the 2017 outburst did not allow for an unambiguous identification of the nature of the compact object. 
MAXI~J1807+132 was detected in outburst again in 2019 and was monitored regularly with  \textit{NICER}.
In this paper we report on five days of observations during which we detected three thermonuclear (Type-I) X-ray bursts, identifying the system as a neutron star LMXB. 
Time-resolved spectroscopy of the three Type-I bursts revealed typical characteristics expected for these phenomena. 
All three Type-I bursts show slow rises and long decays, indicative of mixed H/He fuel. We find no strong evidence that any of the Type-I bursts reached the Eddington Luminosity; however, under the assumption that the brightest X-ray burst underwent photospheric radius expansion, we estimate a $<12.4$\,kpc upper limit for the distance. 
We searched for burst oscillations during the Type-I bursts from MAXI~J1807+132 and found none ($<10\%$ amplitude upper limit at 95\% confidence level).
Finally, we found that the brightest Type-I burst shows a \csim1.6\,sec pause during the rise. This pause is similar to one recently found with \textit{NICER} in a bright Type-I burst from the accreting millisecond X-ray pulsar SAX~J1808.4--3658. The fact that Type-I bursts from both sources can show this type of pause suggests that the origin of the pauses is independent of the composition of the burning fuel, the peak luminosity of the Type-I bursts, or whether the NS is an X-ray pulsar.

\end{abstract}

\begin{keywords}
stars: neutron -- stars: individual (MAXI~J1807+132) -- X-rays: binaries -- X-rays: bursts\vspace{-1em}
\end{keywords}



\section{Introduction}

Low-mass X-ray binaries (LMXBs) consist of either a neutron star (NS) or a black hole (BH) primary accreting material from a low mass ($\lesssim 1$\,M$_{\odot}$) companion star through Roche lobe overflow. Gas flowing from the companion star forms an accretion disc around the primary. As the gas in the disc spirals closer to the compact object, gravitational potential energy is released in the form of X-rays. A LMXB spends most of its life in a quiescent (very low or non-accreting) state, but when in outburst, it can reach X-ray luminosities of $L_{\rm X} \simeq 10^{34-38}$\,erg\,s$^{-1}$ \citep[see, e.g.,][for a review]{tauris2006}.

It is challenging to differentiate between LMXBs hosting a BH or NS.
Dynamical measurements of a compact object's mass can be used to help identify its nature, however, the sensitivity of the observations needed to make these estimations is difficult to acquire \citep[e.g.,][]{cas14}. 
The detection of coherent pulsations confirms a NS identification, as these are associated with the spin frequency of the NS \citep[e.g.,][]{pat12}.
Similarly, thermonuclear burning requires a solid surface for the fuel to settle on. A BH has no solid surface, but an event horizon instead, hence the detection of thermonuclear burning also secures a NS identification. 
There are additional observables that can be used to identify the nature of the compact object, although the identification is based on empirical evidence that certain phenomenology has only been observed in either a BH or a NS so far. 
For example, quasi-periodic oscillations (QPOs) with frequencies higher than 500\,Hz have only been seen in X-ray light curves from observations of NS systems \citep[see, e.g.,][for a review]{vdk06}.
However, the observation and characterisation of QPOs, and the power-spectral broadband noise, with frequencies below a few hundred Hertz are not always conclusive \citep[e.g.,][]{kle08}.
Multi-wavelength observations have been also used to distinguish between BHs and NSs. For example,  NSs can be \csim30 times fainter in the radio band when compared with BHs observed at similar X-ray luminosities. However, recent works have shown that there is a population of radio-faint BHs that have similar radio luminosities to NSs \citep{tet18}.

Thermonuclear burning in a NS atmosphere can manifest as stable, unstable, or marginally stable \citep[e.g.,][]{gal17}. Here, we concentrate on unstable thermonuclear (Type-I) X-ray bursts (hereafter referred to as ``X-ray bursts"). X-ray bursts appear as a sudden increase in X-ray emission over timescales of seconds. They occur when pure or mixed material - Hydrogen, Helium, and sometimes Carbon - accreted onto the NS surface reaches a critical density and temperature which allows for runaway thermonuclear burning \citep[see, e.g.,][for reviews]{lew93, str03}. 
During this process, the X-ray flux increases rapidly ($\lesssim 1-10$\,s) and is generally followed by an exponential-like decay over tens of seconds as the NS atmosphere cools. 

When a Type-I X-ray burst reaches the Eddington limit, $L_{\rm Edd}$, it results in photospheric radius expansion \citep[PRE, see, e.g.,][]{taw84,lew84}, where the outward radiation pressure exceeds the inward gravitational force. Since black body luminosity scales as $L\propto R^2 T^4$, when the X-ray burst reaches the Eddington limit an increase in radius and decrease in temperature can be seen in time-resolved spectral analysis of PRE X-ray bursts. An X-ray burst's luminosity remains roughly constant at $L_{\rm Edd}$ during PRE, thus PRE X-ray bursts can be used as empirical standard candles \citep{van78, kuu03}.

MAXI~J1807+132 (hereafter MAXI~J1807) was first discovered by the nova-alert system of the Monitor of All-sky X-ray Image Gas Slit Camera  \citep[\textit{MAXI}/GSC;][]{mat09,mih11} during its 2017 outburst \citep{2017ATel10208}. This detection was followed up with The Neil Gehrels Swift Observatory \citep[\textit{Swift};][]{geh04} X-Ray Telescope \citep[XRT;][]{bur03} observations, spectral analysis of which suggested a LMXB with a NS primary \citep{shi17}. Further X-ray studies with \textit{XMM-Newton}'s European Photon Imaging Camera \citep[EPIC;][]{xmm} and ground-based optical telescopes supported the NS identification, although the possibility of a BH primary could not be ruled out \citep{jim19}.

After roughly 2 years in quiescence, a new outburst of MAXI~J1807 was detected on 10 September 2019 (MJD 58736) by the \textit{MAXI}/GSC nova-alert system \citep{2019ATel13097}. Subsequently, the Neutron Star Interior Composition Explorer \citep[\textit{NICER;}][]{nicer} monitored the source from 16 September (MJD 58742), and observed the system on a regular basis. The outburst of MAXI~J1807 was characterised by flaring events lasting days to weeks \citep{2019ATel13173}. Two X-ray bursts were detected by \textit{NICER} on 28 and 29 October, and preliminary analysis of these two events reported by \cite{2019ATel13239} confirmed MAXI~J1807 as a neutron star LMXB. 
 In this paper we report on the detection of a third X-ray burst with \textit{NICER}, and present a detailed analysis of all three Type-I bursts.


\section{Observations and Data Analysis}

 \begin{figure}
 \includegraphics[width=\linewidth, trim = 0 0 0 70, clip]{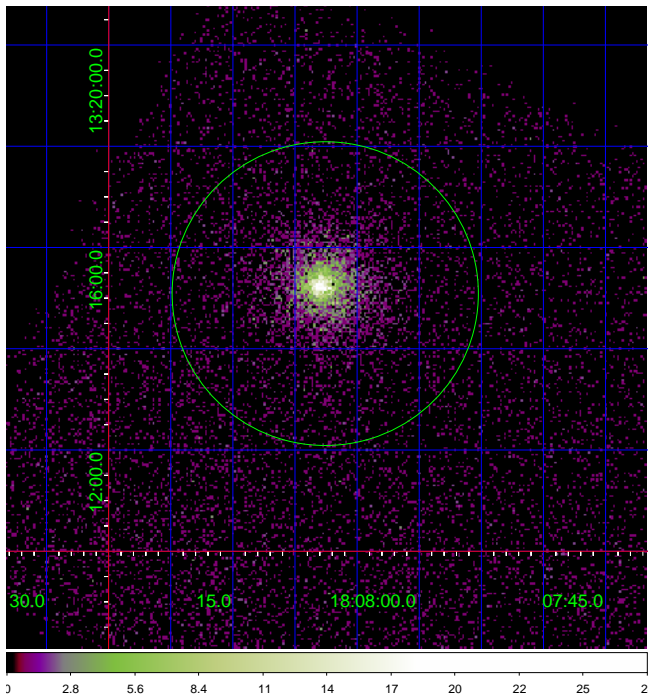}
 \caption{A 3--78\,keV \textit{NuSTAR} image of the \textit{NICER} field-of-view, 26 September 2019 (ObsID 90501342002, exposure time of 22\,ks). The \textit{NICER} field-of-view for our observations is denoted by a green circle of 3\,arcmin radius.} 
 \label{fig:nu}
 \end{figure}

\begin{table}
\setlength{\tabcolsep}{3.6pt}
\centering
 \caption{Details of the \textit{NICER} observations analysed in this paper. Quoted exposure times are as after processing.}
 \label{tab:ob}
 \begin{tabular}{cccccc}
  \hline
  \# & ObsID & Start Time & Date & Exposure & X-ray  \\
   &  &\scriptsize{(MJD)} & \scriptsize{(DD-MM-YY)} & \scriptsize{(sec)} & Burst \\
  \hline
  1 & 2200840122 & 58783.04946 & 27-10-2019 & 15838 & - \\
  2 & 2200840123 & 58784.01681 & 28-10-2019 & 12557 & Yes \\
  3 & 2200840124 & 58785.04902 & 29-10-2019 & 5839 & Yes \\
  4 & 2200840125 & 58786.27471 & 30-10-2019 & 4480 & Yes \\
  5 & 2200840126 & 58787.04898 & 31-10-2019 & 4850 & - \\
  \hline
 \end{tabular}
\end{table}

\textit{NICER} is an X-ray telescope on board the International Space Station. It was launched in 2017 with instrumentation specifically designed for the study of NSs. \textit{NICER}'s X-ray Timing Instrument (XTI) operates in the $0.2-12$\,keV energy band, providing high timing and spectral resolution of 100\,ns and $6 < E/ \Delta E < 80$ from 0.5\,keV to 8\,keV respectively. With 52 active detectors, \textit{NICER} provides an effective area of 1900\,cm$^2$ at 1.5\,keV.

\textit{NICER} observed MAXI~J1807 between 16 September and 26 November 2019, generating a total of 47 observation IDs (ObsIDs). 
We searched all available data for X-ray bursts;
here we report on the five observations around the time of
the detection of three X-ray bursts (see Table \ref{tab:ob}). The observations
correspond to ObsIDs 2200840122--26, performed between
27 October (MJD 58783) and 31 October (MJD 58787), where X-ray bursts are observed in ObsIDs 2200840123--25.
The detailed analysis of the spectral and variability evolution of the full outburst will be presented elsewhere.

To process the data, we used \texttt{HEASOFT} v6.26.1 and \texttt{NICERDAS} v6 \citep{heasoft}, and applied standard filtering criteria, i.e., pointing offset $<$ 54$^{\prime\prime}$, Earth limb elevation angle $>$ 15$^{\circ}$, and bright Earth limb angle $>$ 30$^{\circ}$. 
 Because the observations occurred during an epoch of high optical loading, the parameter ``\texttt{underonly\_range}" was increased to $0-400$ when reprocessing the data using the \texttt{nicerl2} tool. 
 The first X-ray burst occurred during a nominal South Atlantic Anomaly (SAA) passage. As such, ``\texttt{nicersaafilt}" was set to ``no'' on the second observation analysed, in order to include the data taken during the SAA passage. To test if the data were affected by the apparent passage through the SAA, we extracted the $13-15$\,keV lightcurve to look for the presence of high-energy background flares \citep[see][]{bul18}; we found none. The total good exposure after processing was 43.6\,ksec.
 
We determined the background contribution of our observations from \textit{NICER} observations of a Rossi X-ray Timing Explorer \citep[\textit{RXTE};][]{rxte, jah06} blank-sky region (\csim1--2\cps\ from \textit{RXTE}-6).

\subsection{Light Curves and Hardness Ratios}
  To study the X-ray bursts in the context of MAXI~J1807's outburst, we constructed a 0.3--10\,keV long-term light curve using 25\,s bins. 
 We extracted individual X-ray burst light curves in the 0.3--10\,keV energy band using 0.1\,s bins. 
 
 We define the start of each X-ray burst as the first 0.1\,s bin which has an intensity $1\sigma$ above the persistent emission (calculated as the median count rate of the exposure) when scanning backwards in time from the X-ray burst peak.
 We define the end of an X-ray burst as the initial time of a 10\,s bin that is consistent within $1\sigma$ of the median count rate of the last 100\,s of the data segment containing the X-ray burst.

The hardness ratio was defined as the count rate in the 1--10\,keV energy band divided by the count rate in the 0.3--1\,keV energy band. Light curves for the hardness ratios were extracted with 0.2\,s binning.

\subsection{Spectral Analysis}

\begin{table}
\setlength{\tabcolsep}{3.6pt}
\centering
 \caption{The black body temperatures, photon indices, 0.5--10\,keV unabsorbed fluxes, and reduced $\chi^2$ obtained from modelling the pre-burst emissions of MAXI~J1807, with all errors quoted at a 90\% confidence interval.}
 \label{tab:pe}
 \begin{tabular}{ccccc}
  \hline
  Burst & $kT_{bb}$ & Photon  & Unabsorbed Flux & $\chi^2_\nu$ \\
   & \scriptsize{(keV)} &Index &\scriptsize{(10$^{-10}$\,erg\,cm$^{-2}$\,s$^{-1}$)} & \scriptsize{($\chi^2/\mathrm{dof}$)}\\
  \hline
  1 & $0.10\,\pm\,0.01$ & $2.1\,\pm\,0.2$ & $0.54\,\pm\,0.03$ & 205.8/177\\
  2 & $0.19\,\pm\,0.01$ & $2.0\,\pm\,0.2$ & $1.75\,\pm\,0.05$ & 263.2/233\\
  3 & $0.11\,\pm\,0.02$ & $2.6\,\pm\,0.3$ & $1.14\,\pm\,0.04$ & 214.8/178\\
  \hline
 \end{tabular}
\end{table}

\subsubsection{Persistent Emission}\label{sec:pers}

Before examining the X-ray burst emissions, the pre-burst (persistent) spectra of MAXI~J1807 were first explored. For this,
we extracted the spectra of the persistent emission from exposures of 509\,s, 395\,s, and 110\,s before the first, second, and third X-ray bursts, respectively. The 0.5--10\,keV energy spectrum was then successfully fitted with an absorbed black body plus power-law model in \texttt{Xspec} v12.10.1:

\vspace{0.5em}
\noindent \hspace{0.5em}\texttt{tbabs \ (bbodyrad + powerlaw)} ,
\vspace{0.5em}

\noindent where \texttt{tbabs} takes into account the effect of the interstellar absorption \citep{wil00}.
This simple model fits the data well. As the burst flux is generally much brighter than the persistent emission, our results remain unaffected if we use more complex models for the persistent emission.
We found a column density of $N_H = (1.3\,\pm\,0.9) \times10^{21}$\,cm$^{-2}$ before the first X-ray burst and chose the same value to fit the persistent emission spectra before the second and third X-ray bursts. 
We note that our results were the same within error  by (i) using the column density as explained above, (ii) using  column densities as estimated from the persistent emission before each burst, or  (iii) using an average of the three column densities.

The black body temperatures, photon indices, 0.5--10\,keV unabsorbed fluxes, and reduced $\chi^2$ obtained from modelling the pre-burst emissions are reported in Table \ref{tab:pe}.
In all cases, we obtained a reduced $\chi^2$ of \csim1.1. 
The errors are quoted for a 90\% confidence interval.

\subsubsection{Time-Resolved Spectroscopy}

We performed time-resolved spectroscopy on all three X-ray bursts.
Each X-ray burst was divided into time bins containing a minimum of 500 counts. We fitted energy spectra to each time bin using the variable persistent flux method \citep[$f_a$-method; see, e.g.,][]{wor13, wor15}.

 The $f_a$-method describes the X-ray burst thermal emission with a black body component and any possible excess is accounted for by scaling the pre-burst emission. This model is usually constructed by first fixing the pre-burst spectral components and then adding a black body component for the X-ray burst emission. A multiplicative factor $f_a$ is then applied on the persistent part, which allows us to account for the effect of X-ray burst emission on the accretion processes. In \texttt{Xspec}, this model reads as:

\vspace{0.5em}
\noindent \hspace{0.5em}\texttt{tbabs  (\ bbodyrad \ + \ $\mathtt{f_{a}}$ (bbodyrad + powerlaw)\ )} ,
\vspace{0.5em}

\noindent where \texttt{tbabs} takes into account the interstellar absorption and  \texttt{bbodyrad} the black body emission from the X-ray burst. 
In our study, the model for the persistent emission (\texttt{bbodyrad + powerlaw}) is fixed to the respective values we found in Section~\ref{sec:pers}. 

The $f_a$-method has already been used for time-resolved spectroscopy of X-ray bursts using \textit{NICER} data \citep[see, e.g.,][]{kee18, jai19}. Although the $f_a$-method has been shown to be a good way to parameterise the effects on the accretion disc, it is likely that $f_a$ should be a function of energy rather than a constant \citep[see, e.g.,][]{kee14, deg18}. In our analysis, given the relatively low count rates during MAXI~J1807's X-ray bursts, we assumed an $f_a$ that does not depend on energy.


\section{Results and Discussion}

\subsection{X-ray Imaging Observations}

\textit{NICER} is a non-imaging instrument. Here we investigate whether the X-ray bursts could originate from a different source than MAXI~J1807.
 
MAXI~J1807 was observed with \textit{NuSTAR} \citep{nustar} on 26 September 2019, i.e., about 35 days before the first X-ray burst we detected with \textit{NICER}. In Figure \ref{fig:nu} we show the 3--78\,keV \textit{NuSTAR}  image including the \textit{NICER} field-of-view \citep[a circle with 3\,arcmin radius, corresponding to a 30\,arcmin$^2$ field-of-view;][]{nicer}. 
Only one source was significantly detected, with coordinates consistent with that of MAXI~J1807. 
 
There are also 28 photon counting (PC) mode \textit{Swift}/XRT observations taken over the period 26 March--30 May 2017 (there are no \textit{Swift}/XRT imaging observations during 2019). We created a 0.3--10\,keV image of the \textit{NICER} field-of-view by integrating the 28 images (total exposure time of 24\,ks). The image was reduced using the \textit{Swift}/XRT data product generator provided by the University of Leicester\footnote{\url{https://www.swift.ac.uk/user_objects/index.php}}. 
Only one source was significantly detected, with coordinates consistent with that of MAXI~J1807. Given the evidence provided by \textit{NuSTAR} and \textit{Swift}/XRT, we concluded that the X-ray bursts originate from MAXI~J1807.

\subsection{Outburst Evolution and Occurrence of X-ray bursts}
\label{sec:outburst}

\begin{figure}	
    \centering
    \includegraphics[width=\linewidth, trim=0 10 0 5, clip]{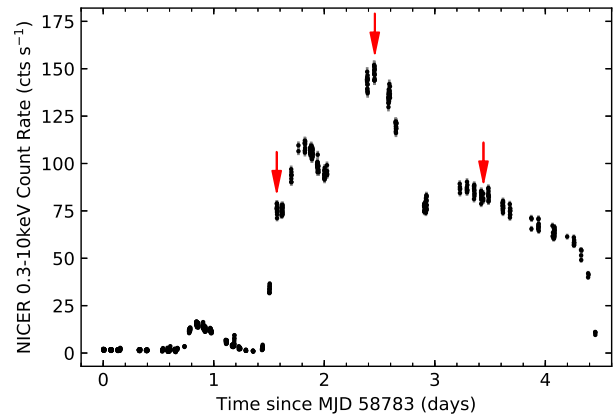}
    \caption{\textit{NICER} long-term light curve of MAXI~J1807 with 25\,s time resolution in the 0.3--10\,keV energy band. This section of the outburst contains the three thermonuclear X-ray bursts detected in October 2019. The X-ray burst data have been removed for clarity, and their onsets marked by arrows.
    }
    \label{fig:outburst}
\end{figure}

\begin{figure*}
    \centering
    \includegraphics[width=\linewidth, trim=60 20 60 40, clip
    ]{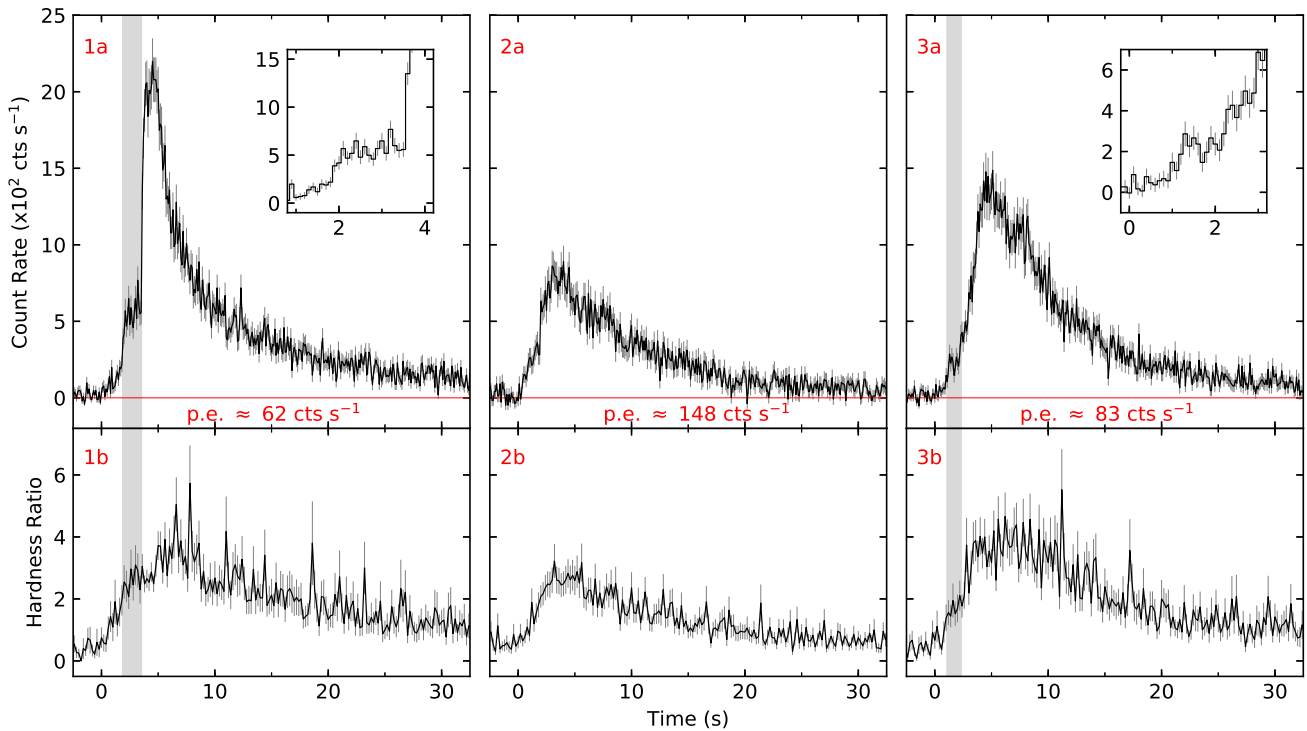}
    \caption{Top panels: light curves of the three X-ray bursts from MAXI~J1807 with 0.1\,s time resolution in the 0.3--10\,keV energy band. We found the persistent emission levels by taking the mean count rate for 100\,s of data before the start of each X-ray burst, and subtracted them from the light curves. The persistent emission count rates are denoted on the panels as ``p.e.". The insets in panels 1a and 3a show rise features in detail, which are also shaded in the main panels. Bottom panels: hardness ratios (1--10\,keV / 0.3--1\,keV count rate) with 0.2\,s time resolution.
     }
    \label{fig:bursts}
\end{figure*}

Figure \ref{fig:outburst} shows the outburst evolution of MAXI~J1807 from 27 to 31 October 2019 (MJD 58783--58787). The occurrence of the X-ray bursts are marked by arrows.  
On 27 October ($t\approx0.7$\,days), the light curve shows a flare-like feature, where the count rate rises from \csim2\cps\ to \csim15\cps\ and falls back to the original count rate within half a day. Roughly 3\,hrs later, we observe the onset of a larger flare with multiple peaks. The count rate increases to \csim70\cps\ over approximately 3\,hrs, after which the first X-ray burst occurs on MJD 58784.61727 ($t\approx1.6$\,days in Figure \ref{fig:outburst}). The flare reaches a first peak at \csim110\cps\  roughly 4.5\,hrs after the X-ray burst, and drops to \csim95\cps\ in the following 4\,hrs. There is an additional peak  almost 10\,hrs later during which the second X-ray burst occurs on MJD 58785.50558 ($t\approx2.5$\,days in Figure \ref{fig:outburst}) at a persistent count rate of \csim150\cps. The persistent count rate then drops to \csim80\cps\ over approximately 10\,hrs before a final rise to \csim90\cps, after which the flare begins to decay. At this stage in the outburst the third X-ray burst occurs on MJD 58786.48803 ($t\approx3.5$\,days in Figure \ref{fig:outburst}) at a persistent count rate of \csim80\cps. There is a sudden decrease in flux just less than a day after the third X-ray burst, where we observe the flux decreasing from \csim60\cps\ to \csim10\cps\ in roughly 4\,hrs.

We observed the three X-ray bursts over a period of three days. The waiting time between the first and second X-ray bursts was 21.3\,hrs, while between the second and third X-ray bursts was 23.6\,hrs. We note that the data-gaps prevent us to understand whether there were other X-ray bursts than those we detected.

\subsection{Type-I X-ray Burst Light Curves} 
\label{sec:bursts}

Figure \ref{fig:bursts} shows light curves and hardness ratios of the three X-ray bursts exhibited by MAXI~J1807. The first, second, and third X-ray bursts will hereafter be referred to as B1, B2, and B3, respectively. 
At the burst onset, $t_{B1}=0$, B1 count rate increases from the persistent rate of \csim62\cps\ to \csim500\cps\ over approximately 2\,s. At this point, the burst exhibits a ``pause'' lasting \csim1.6\,s (shaded region and inset in Figure \ref{fig:bursts}, panel 1a). Following the pause, the burst reaches its peak in roughly 0.2\,s. The count rate remains approximately constant during this peak at \csim2250\cps\ between $t_{B1} = 3.8 - 5.1$\,s. After the peak, the count rate decreases for roughly 133\,s to a persistent rate of \csim70\cps, i.e., at a flux \csim13\% higher than the persistent emission before the burst onset. 
The overall burst duration is \csim137\,s.

The rise of B2 is in two parts. In the initial rise, starting at $t_{B2}=0$, the count rate increases from the persistent rate of \csim148\cps\ to \csim460\cps\ over approximately 2\,s. 
The count rate then rises suddenly and reaches the peak of the burst in less than one second. This peak has an average count rate of \csim920\cps\ between $t_{B2} = 2.6 - 4.6$\,s. 
As the data segment containing the burst ends less than 100\,s after the burst peak, the end of the burst is ill-defined while the count rate is still 8\% above the pre-burst level (160\cps, the median count rate of the last 20\,s of the data segment containing B2).

B3 exhibits interesting features in both its rise and decay. 
At $t_{B3}=0$, B3's count rate increases for roughly 1.5\,s from the persistent rate of \csim83\cps\ to a potential pause similar to that we observed in B1, but at \csim290\cps\ (marked by the shaded region and inset in Figure \ref{fig:bursts}, panel 3a). 
After the potential pause, the count rate continues to rise reaching a peak average count rate of \csim1417\cps\ between $t_{B3} = 4.0 - 5.3$\,s. During the decay there is indication of a double peak at $t \approx 7.5$\,s lasting roughly 2\,s; however the double peak is not statistically significant given the large error bars. After this, the count rate decreases over approximately 78\,s to a post-burst persistent count rate of \csim100\cps, which is \csim20\% higher than the pre-burst persistent rate. The overall burst duration is \csim83\,s.

The hardness ratios of all three X-ray bursts track similar profiles to the light curves, increasing through the burst rise and decreasing through the decay. The hard band (1--10\,keV) dominates after less than 1\,s in each case, and increases to peaks of \csim4 in B1 and B3, and \csim3 in B2. There is a plateau in the hardness ratio that starts during the pause in B1 and appears to continue after it. There is also evidence of a plateau during the potential pause found in B3, after which the ratio increases; however in this case the plateau is not as well constrained as in B1.  

We measure different average count rates for the persistent emission before and after each X-ray burst. Given our data set, we are unable to understand if this is due to an intrinsic change in the flux of the persistent emission, or whether it is the effect of long decay tails which we cannot differentiate from the continuum \citep[see, e.g.,][]{int17}.
Out of the three X-ray bursts, B1 has the highest peak count rate, whilst B2 has the lowest. All three X-ray bursts have a rise time of \csim4\,s and exhibit long decay tails (>1 minute). A slow rise and long decay is indicative of H-rich fuel at the moment of ignition, which is likely the result of accretion of a mixed H/He fuel \citep{gal08, gal17, sch01}.

\subsection{X-ray Burst Spectral Characteristics}

\begin{figure*}
    \centering
    \includegraphics[width=\linewidth, trim=75 30 80 40, clip]{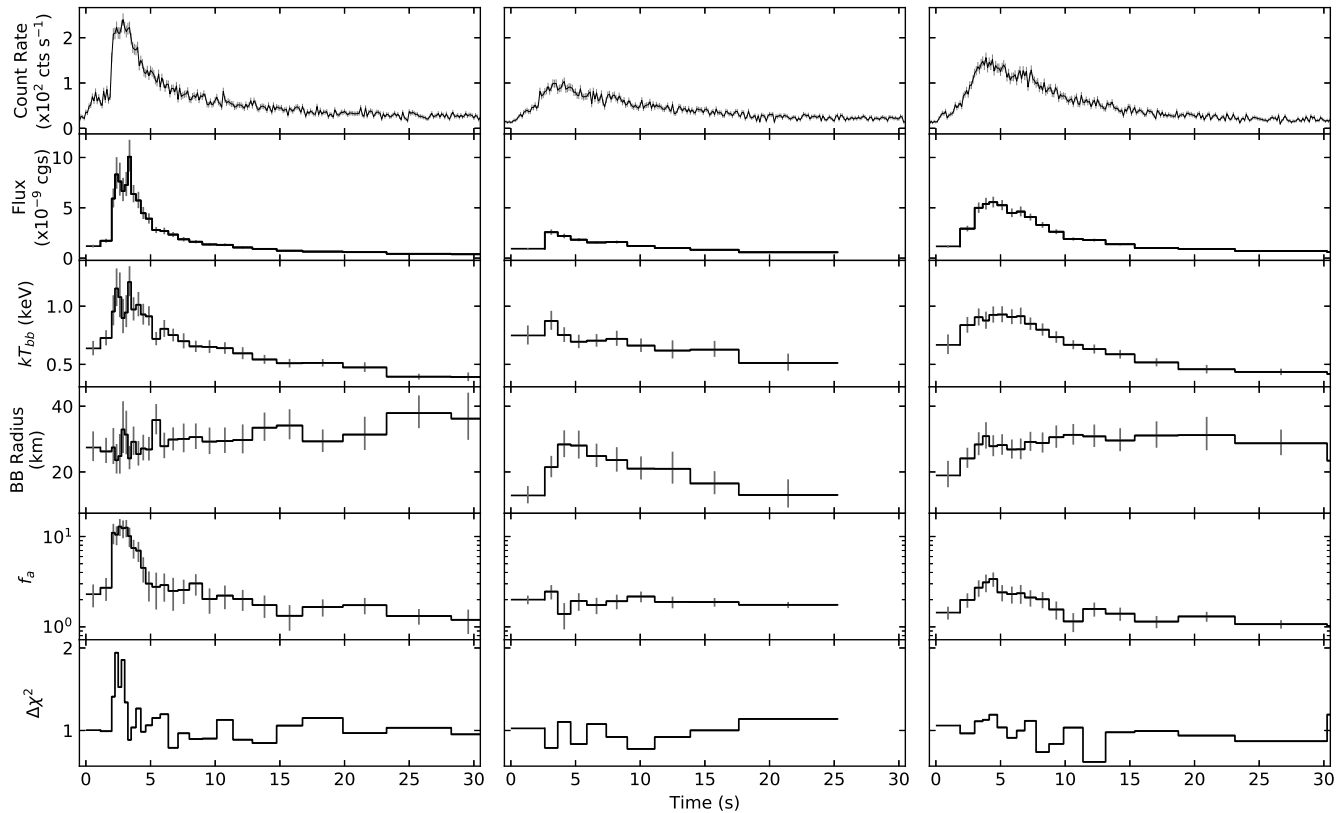}
    \caption{Time-resolved spectroscopy of the three X-ray bursts (B1, B2 and B3, left to right, respectively) using the variable $f_a$ method. From top, we show the count rate, the estimated bolometric flux (in units of erg\,s$^{-1}$\,cm$^{-2}$), the black body temperature and radius (assuming a distance of 12.4 kpc), the variable scale factor $f_a$ and the reduced $\chi^2$. 
    }
    \label{fig:spec}
\end{figure*}

In Figure \ref{fig:spec} we show the best-fit parameters for the  time-resolved spectroscopy of the X-ray bursts.
The bolometric unabsorbed flux ($F_{\mathrm{bol}}$) of all three X-ray bursts approximately follows the light curve contours. The peak fluxes are reported in Table \ref{tab:bb}. 
The black body temperature ($kT_{bb}$) also follows the contours of the burst profiles in the light curves, increasing during the burst rise and decreasing during the decay.  
The evolution of $kT_{bb}$ is typical of the heating and cooling processes of the NS atmosphere during a Type-I X-ray burst \citep[see, e.g.,][for a review]{lew93}, thus strongly suggesting that these features are indeed thermonuclear in origin. 
The peak temperatures achieved by each of MAXI~J1807's X-ray bursts (\csim 1\,keV, reported in Table \ref{tab:bb}) are lower than what has typically been seen in previous works for other sources ($\gtrsim2$\,keV).
We did not find any evidence that the low temperatures are due to any instrumental effects or due to our spectral modelling. Indeed, previous results on other sources using \textit{NICER} data and the $f_a$-method have shown that X-ray bursts can reach temperatures in the 2--3\,keV range \citep[see, e.g.,][]{kee18}. Therefore, we conclude that the low temperatures we measure are intrinsic to the X-ray bursts detected in this source.

Thermonuclear X-ray bursts can be used to estimate the distance to the system if they exhibit photospheric radius expansion. The brightest of the three X-ray bursts, B1, does not show strong evidence, if any, of PRE. However, under the assumption that B1 reached the Eddington luminosity ($L_{\rm Edd}$), MAXI~J1807 contains a 1.4\,$M_{\odot}$ NS and $L_{\rm Edd}=2\times10^{38}$\,erg\,s$^{-1}$ \citep[as expected for H-rich fuel, see, e.g.,][]{lew93}, we found a distance upper limit of 12.4\,kpc.

We see differences in the evolution of the black body radius, calculated assuming a distance of 12.4\,kpc, during the three X-ray bursts. For B1, it remains approximately constant at \csim28.5\,km. The radius during B2 approximately follows the contour of the light curve, peaking at \csim28.4\,km. In the case of B3, the radius increases throughout the rise in count rate, and then remains roughly constant at \csim28.5\,km after the burst peak.
The radii reported here are larger than what might be expected \citep[see, e.g.,][]{gal08}. This is due to the fact that they are calculated using the 12.4\,kpc upper limit for the distance. If MAXI~J1807 were at, for example, half this distance, the peak radius would be roughly 18.5\,km, which, whilst it is still large, is more consistent with the expected radius during X-ray bursts.

The parameter $f_a$ describes the photospheric evolution of the X-ray burst as well as its effect on the accretion processes.
For B2, $f_a$ remains at \csim1.9 throughout the burst.
For B1 and B3, $f_a$ increases and peaks at the same time as the burst, and decreases afterwards. This can be the result of real changes in the accretion disc or due to the choice of model.  Testing whether the accretion disc varies significantly is difficult to assess \citep[e.g.,][]{deg18}, particularly with the low count rates in our data. 

In Table \ref{tab:bb}, we report the X-ray burst fluences and time scales $\tau$ ($=$ fluence / peak $F_{bol}$). We calculated the fluences of B1, B2, and B3 over the first 31\,s, 25\,s, and 42\,s of data, respectively (during these time-intervals the fluence was well constrainted). The fluences are in the (\csim2 to \csim3.5)\,$\times 10^{-8}$ erg\,cm$^{-2}$ range, which is the lower end of typical fluence distributions found in other sources \citep[][]{gal08, lyu15, boi07}. However, given the uncertainties in our calculations of X-ray burst duration and the intensity of the continuum after the burst, our fluences are probably are underestimated by 10--15\%. 
We found that $\tau$ was in the \csim2 to \csim6\,s range. These values are also in the lower end of the distribution usually found in other sources \cite{gal08,gal20}.

Calculating an accurate value for the X-ray burst parameter $\alpha$ \citep[ratio of the integrated persistent flux to the burst fluence, see, e.g.,][]{gal08} is not possible without a well-constrained recurrence time. However, as an exercise, we calculate a rough estimate for $\alpha$ with certain caveats. We only detected X-ray bursts from MAXI~J1807's 2019 outburst  during a \csim1.9\,day window. During this window,  \textit{NICER} covered \csim8\% of the time.
If we assume that only \csim8\% of the X-ray bursts were detected in this time frame, this implies a total of \csim36 X-ray bursts with a recurrence time of \csim1.27\,hrs. Taking a pre-burst flux of \csim1$\times 10^{-10}$\,erg\,s$^{-1}$\,cm$^{-2}$ and fluence of \csim2$\times 10^{-8}$\,erg\,cm$^{-2}$, this gives an averaged $\alpha\approx23$. These assumptions make the statistical and systematic uncertainties on this calculation rather large. Comparing this value with what we see in \cite{gal08}, our estimate for $\alpha$ is much smaller than what we would expect for our calculated values of $\tau$ \citep[if compared with other sources, we expect $\alpha > 60$ for $\tau < 10$; see Figure 14,][]{gal08}. The uncertainty in our calculations may be able to account for this tension, although it may also be an intrinsic property of the X-ray bursts in MAXI~J1807 as the fluences we measured are smaller than average.

\begin{table}
\setlength{\tabcolsep}{2pt}
\centering
 \caption{The best fit parameters obtained from time-resolved spectroscopy for peak black body temperatures and bolometric fluxes achieved by each X-ray burst from MAXI~J1807, and calculated estimates for X-ray burst parameters fluence and $\tau$.}
 \label{tab:bb}
 \begin{tabular}{ccccc}
  \hline
  Burst & Peak $kT_{bb}$ & Peak $F_{\mathrm{bol}}$ & Fluence & $\tau$\\
   & \scriptsize{(keV)} & \scriptsize{($10^{-8}$ erg\,s$^{-1}$\,cm$^{-2}$)} & \scriptsize{($10^{-8}$ erg\,cm$^{-2}$)} & \scriptsize{(s)}\\
  \hline\noalign{\vspace{1mm}}
  1 & $1.21^{+0.14}_{-0.12}$    & $1.01^{+0.17}_{-0.15}$    & $2.30\pm0.29$     &  $2.29\pm0.07$ \\\noalign{\vspace{0.5mm}}
  2 & $0.9\pm0.1$               & $0.26\pm0.03$             & $1.36\pm0.11$     &  $5.30\pm0.13$ \\\noalign{\vspace{0.5mm}}
  3 & $0.9\pm0.07$              & $0.56^{+0.06}_{-0.05}$    & $3.17\pm0.27$     &  $5.70\pm0.06$ \\\noalign{\vspace{0.5mm}}
  \hline
 \end{tabular}
\end{table}

\subsection{Search for Burst Oscillations}

To search for burst oscillations, we constructed 0.8--8.0\,keV X-ray
burst light curves at 1/8192\,s time resolution. Each light curve
started 10\,s prior to the burst onset, and had a duration of 50\,s, such that the entire burst epoch was included. We then applied
a $T=2$\,s, 4\,s, and 8\,s wide window to each light curve, which we moved across the respective burst in steps of $T/2$. For each combination of
window size and position we constructed the power spectrum and searched
between 100\,Hz and 1000\,Hz for the presence of a coherent burst
oscillation signal in excess of the counting noise distribution (see \citealt{wat05}; and \citealt{wat12} for a review). No significant signals were observed to a 95\% confidence upper limit of approximately 10\% fractional amplitude at peak intensity.

\subsection{Pauses and Comparison}

Historically, the rises of thermonuclear Type-I X-ray bursts have generally been described as relatively fast and smooth. In the last two decades, thanks to the high time resolution and effective area of \textit{RXTE}, we have learned that X-ray burst profiles can be very complex. 
These include slower or faster rises, two-step rises (like we see in panel 2a of Fig \ref{fig:bursts}), concave and convex rises \citep{mau08}, and X-ray bursts with multiple (2 or 3) peaks. We note that, sometimes an X-ray burst with a double-peaked profile can be interpreted as two different events \citep[e.g., ][]{bha07}.

X-ray bursts with multiple peaks in their bolometric luminosity profile are not uncommon \citep[see, e.g.,][and references therein]{gal08, lew93, wat07}. Notably, the source 4U~1636--536 shows the largest sample of multi-peaked X-ray bursts with a variety of profiles, including X-ray bursts with two peaks that reach the same luminosity \citep[e.g.,][]{bha06}, X-ray bursts where the two peaks are not at the same luminosity \citep[e.g.,][]{bha06b, mau08}, and even triple-peaked X-ray bursts \citep{van86, zha09}. 
Double-peaked X-ray bursts from this source have been interpreted in a number of different ways. For example, a second peak may be caused by multi-step thermonuclear energy release, or the ignition of fresh or leftover material \citep[see, e.g., ][and references therein]{jai19}.
\cite{bha06, bha06b} interpreted the double-peaked profile as the result of a temporary stalling in the burning front at the NS's equator. This interpretation requires rare high-latitude ignition \citep{spi02}, explaining why double-peaked bursts are far less common than single-peaked bursts \citep[e.g.,][but also see discussions by \cite{coo07,wat07,mau08}]{gal08}. 
However, triple-peaked bursts are difficult to explain solely by a temporary stalling in the burning front. \cite{zha09} showed that observations of triple peaks pose particular challenges for the polar ignition mechanism required by the interpretation of \cite{bha06, bha06b}.

In this paper, we report on the discovery of a $<$\,2\,sec pause in the rise of one single-peaked burst in the LMXB MAXI~J1807+132. Given the complex profiles that X-ray bursts may have (see above), it is important to understand whether pauses during an X-ray burst rise are a different manifestation of the multi-peak phenomena (i.e. what we observe as a pause is in reality a very short, weak peak of a double-peaked X-ray burst), or the pause is unrelated to the physical process that produces multi-peak X-ray bursts.

\cite{bul19} recently reported on a bright, He-fueled PRE Type-I X-ray burst from the accreting millisecond X-ray pulsar SAX~J1808.4--3658. The light curve profile of the X-ray burst (hereafter S1) was complex: it showed a pause in the rise and a double peak unrelated to the PRE event. Additionally, S1 showed burst oscillations.
\cite{bul19} found that the pause, the dip between the X-ray burst's double peaks, and the onset of burst oscillations all occur at similar count rates, suggesting that the similar count rates are related to the Eddington limit of a hydrogen envelope.

The pause in the first burst we detected in MAXI~J1807 (B1) is \csim1.6\,s in duration, and occurs at roughly 20\% of the peak count rate. The pause in SAX~J1808.4--3658's S1 is shorter, lasting \csim0.7\,s, and occurs at  roughly 40\% of the peak count rate. We calculated the ratio between peak and pause fluxes to be $5.8^{+0.3}_{-0.2}$, whereas this ratio was calculated to be $1.68\pm0.13$ for S1. 
In both B1 and S1, the rise time before the pause is roughly the same duration as the pause, lasting \csim1.6\,s and \csim0.6\,s for B1 and S1 respectively. However, the rise time from the end of the pause to the burst peak in B1 is \csim0.5\,s,  much shorter than that of \csim3\,s in S1. B1 and S1 also have different fuel compositions, with mixed H/He (H-rich) and He only, respectively.

Although we currently have a sample of only two X-ray bursts with confirmed pauses during the rise, we suggest that the pauses are not a shorter and weaker manifestation of the multi-peaked X-ray bursts seen in other sources.
Our interpretation is supported by the fact that SAX~J1808.4--3658's X-ray burst \citep{bul19} exhibits both a pause and a double-peaked profile, showing that the two phenomena can manifest in the same X-ray burst. In addition, the X-ray burst we report for MAXI~J1807 shows a pause and a single-peaked profile. These two results together, and the lack of other reports on pauses in X-ray bursts studied with \textit{RXTE}, suggest that there is no link between the detection of double-peaked profiles and the detection of a pause during the rise. 
However, our interpretation would be incorrect if the pauses \textit{are} short and weak manifestations of a peak in the X-ray burst profile, meaning that the X-ray burst in SAX~J1808.4--3658 can be considered a triple-peaked X-ray burst, and B1 in MAXI~J1807 can be considered a double-peaked burst.

\cite{int03} reported on an unusual Type-I X-ray burst from GRS~1747-312\footnote{GRS 1747-312 is a transient X-ray source located in the core of the 9.5\,kpc distant globular cluster Terzan 6. Given \textit{RXTE}/Proportional Counter Array's $1^\circ$ field-of-view, \cite{int03} could not discard the possibility that the X-ray burst originated in a different NS-LMXB} observed with \textit{RXTE} whilst the system was apparently in quiescence. The 2--10\,keV light curve profile exhibits a pause-like feature in the rise of the X-ray burst, while the full 2--60\,keV light curve did not show a pause. The pause in the 2--10\,keV light curve is considerably longer than those seen in S1 and B1 (\csim20\,s), and, judging from Figure 2a of \cite{int03}, the pause occurs at roughly 70\% of the peak count rate. 
The X-ray burst also had an unusually strong PRE that was \textit{not} at the start of the X-ray burst, as is expected.  
It is unclear whether this is a longer manifestation of the pause phenomena we see in B1 and S1, but it shows that the presence of pause-like features can depend on the energy band used.

Independently of whether the pauses are related to the multi-peaked phenomena discussed above, our results, together with those presented by \cite{bul19}, show that short pauses in the rise of X-ray burst light curves occur in at least two sources, suggesting that these types of pauses might be present in the X-ray light curves of Type-I bursts from other NS-LMXBs.
We only know of two sources (the accreting millisecond X-ray pulsar SAX~J1808.4--3658 and MAXI~J1807), each with one reported X-ray burst with a clear pause during the rise. This small sample size prevents us from a detailed study on the relation between pauses and X-ray burst characteristics/properties, and naturally it is unclear whether the pauses seen in B1 and S1 are caused by the same mechanism. 
However, under the assumption that we are seeing manifestations of the same phenomena in B1 and S1, even with a sample of two we can conclude that pauses in the X-ray bursts are possible in both pulsating and non-pulsating systems, suggesting that the strength of the magnetic field might not play an important role, unless the lack of pulsations in MAXI~J1807 is due to an alignment between rotation and magnetic axes \citep{lam09,lam09b}. 
In addition, the occurrence of the pause does not appear to depend on the composition of the burning material either, or whether the X-ray burst reaches PRE. 
To our knowledge, standard X-ray burst models do not predict pauses in the X-ray burst rise. However, we note that \cite{kee12} finds X-ray bursts with pauses in their rise when modelling X-ray bursts that occur shortly after a superburst.
In the hot, He-rich atmospheres following a superburst,  X-ray bursts are initially powered by the triple-$\alpha$ process. As the atmospheric temperature begins to rise due to these X-ray bursts, $\alpha$-captures start playing a role powering the peaks of the X-ray bursts. It is during this transition that X-ray bursts with pauses in the rise occur, as it takes some time to ignite the second burning stage through $\alpha$-captures.
At the moment, it is unclear whether the model of \cite{kee12} could be adapted to explain our results, i.e. whether an X-ray burst can show pauses independently of the occurrence of a superburst to create a hot He-rich atmosphere.
%


\section*{Acknowledgements}

A.A. thanks M. Pahari and M. Ozbey-Arabaci for discussion on the X-ray imaging observations. A.A. also thanks the referee for making helpful suggestions that improved the paper, and thanks C. L. Greenwell and N. Wragg for general discussion and support.
D.A. acknowledges support from the Royal Society. G.C.M. and D.A. acknowledge support from the Royal Society International Exchanges ``The first step for High-Energy Astrophysics relations between Argentina and UK".
S.R. acknowledges support from the Chinese Academy of Science (CAS) President's International Fellowship Initiative (PIFI), National Natural Science Foundation of China (NSFC) grant, and the International Postdoctoral Exchange fellowship.
T.G. has been supported in part by the Scientific and Technological Research Council (T\"UBITAK) 119F082, Royal Society Newton Advanced Fellowship, NAF$\backslash$R2$\backslash$180592, and Turkish Republic, Directorate of Presidential Strategy and Budget project, 2016K121370.
C.M. is supported by an appointment to the NASA Postdoctoral Program at the Marshall Space Flight Center, administered by Universities Space Research Association under contract with NASA. 

\section*{Data Availability}

This work made use of data provided by the High Energy Astrophysics Science Archive Research Center (\textsc{heasarc}).





\bibliographystyle{mnras}
\bibliography{ref} 



\appendix



\bsp	
\label{lastpage}

\typeout{get arXiv to do 4 passes: Label(s) may have changed. Rerun}
\end{document}